\documentclass[english,aps,prl,superscriptaddress,twocolumn]{revtex4-1}

\usepackage[T1]{fontenc}
\usepackage{natbib}
\usepackage[latin9]{inputenc}
\usepackage{graphicx}
\usepackage{babel,bm}
\usepackage{amsfonts}
\usepackage{amsmath}
\usepackage{amssymb}
\usepackage{graphicx}
\usepackage{color}
\usepackage{epstopdf}
\usepackage{hyperref}

\usepackage{times}

\newcommand{\ket}[1]{\left| #1 \right\rangle}
\newcommand{\bra}[1]{\left\langle #1 \right|}

\definecolor{Blue}{rgb}{0,0,1}
\definecolor{Red}{rgb}{1,0,0}
\definecolor{Green}{rgb}{0,1,0}
\definecolor{Purp}{rgb}{.2,0,.2}
\definecolor{white}{rgb}{1,1,1}

\begin{document}
\title{Irreversibility and the arrow of time in a quenched quantum system}
\author{T. B. Batalh\~{a}o}
\affiliation{Centro de Ci\^{e}ncias Naturais e Humanas, Universidade Federal do ABC,
Av. dos Estados 5001, 09210-580 Santo Andr\'{e}, S\~{a}o Paulo, Brazil}
\affiliation{Faculty of Physics, University of Vienna, Boltzmangasse 5, Vienna A-1090, Austria}
\author{A. M. Souza}
\affiliation{Centro Brasileiro de Pesquisas F\'{i}sicas, Rua Dr. Xavier Sigaud 150,
22290-180 Rio de Janeiro, Rio de Janeiro, Brazil}
\author{R. S. Sarthour}
\affiliation{Centro Brasileiro de Pesquisas F\'{i}sicas, Rua Dr. Xavier Sigaud 150,
22290-180 Rio de Janeiro, Rio de Janeiro, Brazil}
\author{I. S. Oliveira}
\affiliation{Centro Brasileiro de Pesquisas F\'{i}sicas, Rua Dr. Xavier Sigaud 150,
22290-180 Rio de Janeiro, Rio de Janeiro, Brazil}
\author{M. Paternostro}
\affiliation{Centre for Theoretical Atomic, Molecular and Optical Physics, School
of Mathematics and Physics, Queen\textquoteright{}s University, Belfast
BT7 1NN, United Kingdom}
\author{E. Lutz}
\affiliation{Department of Physics, Friedrich-Alexander-Universit\"at Erlangen-N\"urnberg, 91058 Erlangen, Germany}
\author{R. M. Serra}
\affiliation{Centro de Ci\^{e}ncias Naturais e Humanas, Universidade Federal do ABC,
Av. dos Estados 5001, 09210-580 Santo Andr\'{e}, S\~{a}o Paulo, Brazil}

\definecolor{Blue}{rgb}{0,0,1}
\definecolor{Red}{rgb}{1,0,0}
\definecolor{Green}{rgb}{0,1,0}
\definecolor{Purp}{rgb}{.2,0,.2}

\newcommand{\verify}[1]{{\color{Red} #1}}
\newcommand{\revision}[1]{{\color{Blue} #1}}
\newcommand{\notes}[1]{{\color{Purp} #1}}

\begin{abstract}
%
Irreversibility is one of the most intriguing concepts in physics. While microscopic physical laws are perfectly reversible, macroscopic average behavior has a preferred direction of time. According to the second law of thermodynamics, this arrow of time is associated with a positive mean entropy production. Using a nuclear magnetic resonance setup, we measure the nonequilibrium entropy produced in an isolated spin-1/2 system following fast quenches of an external magnetic field and experimentally demonstrate that it is equal to the entropic distance, expressed by the Kullback-Leibler divergence, between a microscopic process and its time-reverse. 
Our result addresses the concept of irreversibility from a microscopic quantum standpoint. 
\end{abstract}
\maketitle

 
The microscopic laws of classical and quantum mechanics are time symmetric, and hence reversible.  However, paradoxically, macroscopic phenomena are not time-reversal invariant \cite{leb93,zeh07}. This fundamental asymmetry defines a preferred direction of time that is characterized by a mean entropy production. Regardless of the details and nature of the evolution at hand, such entropy production is bound to be positive by the second law of thermodynamics \cite{cal85}. Since its introduction by Eddington  in 1927 \cite{Eddington}, the thermodynamic arrow of time has not been tested experimentally at the level of a quantum system. 

Introduced by Clausius in the form of an uncompensated heat, the importance of  the entropy production in nonequilibrium statistical physics has been recognized by Onsager and  further developed by Meixner, de Groot and Prigogine \cite{vel11}. Defined as $\Sigma= \beta( W- \Delta F)$, for a system at constant inverse temperature $\beta=1/(k_BT) $, where $ W$ is the total work done  on the system and $\Delta F$ the free energy difference ($k_B$ is the Boltzmann constant), it plays  an essential  role in the evaluation of the efficiency of thermal machines, from molecular motors to car engines~\cite{cal85}.

\begin{figure}[b!]
\includegraphics[scale=0.5]{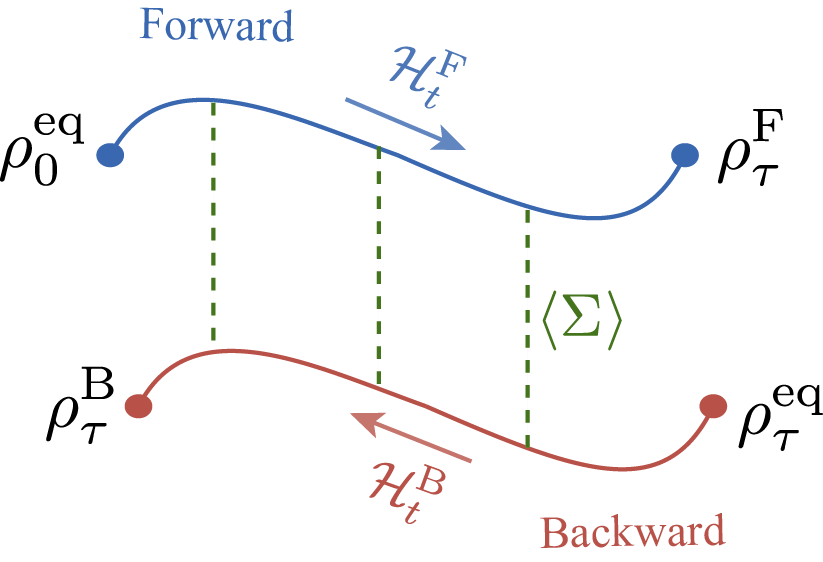}
\caption{ A quantum system (with Hamiltonian ${\cal H}^\text{F}_0$) is initially prepared in a thermal state ${\rho}^\text{eq}_0 $ at inverse temperature $\beta$. It is driven by a fast quench into the nonequilibrium state ${\rho}^\text{F}_\tau $ along a forward protocol described by the Hamiltonian ${\cal H}^\text{F}_t$. In the backward process, the system starts in the equilibrium state ${\rho}^\text{eq}_\tau$ corresponding to the final Hamiltonian ${\cal H}^\text{F}_\tau$ and is driven by the time-reversed Hamiltonian ${\cal H}^\text{B}_t=- {\cal H}^\text{F}_{\tau-t}$ to ${\rho}^\text{B}_\tau $.
The  entropy production $\left\langle \Sigma \right\rangle$ at time $t$ is given  by the Kullback-Leibler divergence between forward and backward states ${\rho}^\text{F}_{t} $ and ${\rho}^\text{B}_{\tau-t}$ [cf. Eq.~\eqref{eq:rel-entropy}].}
\label{fig:process}
\end{figure} 

Starting with Boltzmann's work on the so-called $H$-theorem, the quest for a general microscopic expression for the entropy production, especially far from equilibrium, has been a challenge for more than a century \cite{leb93}. In the last years, formulas for the entropy production and entropy production rate in terms of the microscopic density operator $\rho$ of a system have been obtained for relaxation \cite{sch80}, transport \cite{spo78}, and driven  processes in closed and open quantum systems \cite{def10,def11}. At the same time, the recent development of fluctuation theorems \cite{Esposito,Campisi} has led to a sharpening of the formulation of the second law. Regardless of the size of a system, the arrow of time originates from the combination of an explicit time-dependence of the Hamiltonian of the system and the specific choice of an initial equilibrium state. While the first ingredient breaks time homogeneity (thus inducing the emergence of an arrow of time), the second specifies its direction~\cite{CampisiEntropy}.

 In small systems, thermal and quantum fluctuations are both significant, and fluctuation theorems quantify the occurrence of negative entropy production events during individual processes \cite{jar11}. In particular, the average entropy production $\langle\Sigma\rangle$ for evolution in a time window $\tau$ has been related to the Kullback-Leibler relative entropy between states ${\rho}^{\text{F}}_{t}$ and ${\rho}^{\text{B}}_{\tau-t}$ along the forward and backward (i.e. time reversed) dynamics \cite{Kawai,vai09,par09} (see  Fig.~\ref{fig:process}). Explicitly
\begin{equation}
\label{eq:rel-entropy}
 \langle \Sigma\rangle 
= S\left( {\rho}^\text{F}_{t} \Vert \, {\rho}^{\text{B}}_{\tau-t}\right) = \text{tr} \left[{\rho}^\text{F}_{t} \left(\ln{\rho}^\text{F}_{t} -\ln {\rho}^{\text{B}}_{\tau-t}\right)\right].
\end{equation}
The above equation quantifies irreversibility at the microscopic quantum level and for the most general dynamical process
responsible for the evolution of a driven closed system. A process is thus reversible, $\langle \Sigma\rangle =0$, if forward and backward microscopic dynamics are indistinguishable. Nonequilibrium entropy production and its fluctuations have been measured in various classical systems, ranging from biomolecules \cite{Liphardt} and colloidal particles~\cite{bli06} to levitated nanoparticles~\cite{Gieseler} (see Refs.~\cite{sei12,cil13} for a review). Evidence of time asymmetry has been further observed in a driven classical Brownian particle and its electrical counterpart \cite{and07}. However, quantum experiments have remained  elusive so far, owing to the difficulty to measure thermodynamic quantities in the quantum regime. To date, Eq.~\eqref{eq:rel-entropy} has thus never been tested.

In this Letter, we use a nuclear magnetic resonance (NMR) setup to provide a clear-cut assessment of Eq.~\eqref{eq:rel-entropy} where $ \langle \Sigma\rangle$ and $S\left( {\rho}^\text{F}_{t} \Vert \, {\rho}^{\text{B}}_{\tau-t}\right)$ are tested and evaluated independently. Our methodological approach is founded on the reconstruction of the statistics of work and entropy, following a non equilibrium process implemented on a two-level system, therefore assessing the emergence of irreversibility (and the associated arrow of time) starting from a genuine microscopic scale.

\begin{figure}[b!]
\includegraphics[width=1\columnwidth]{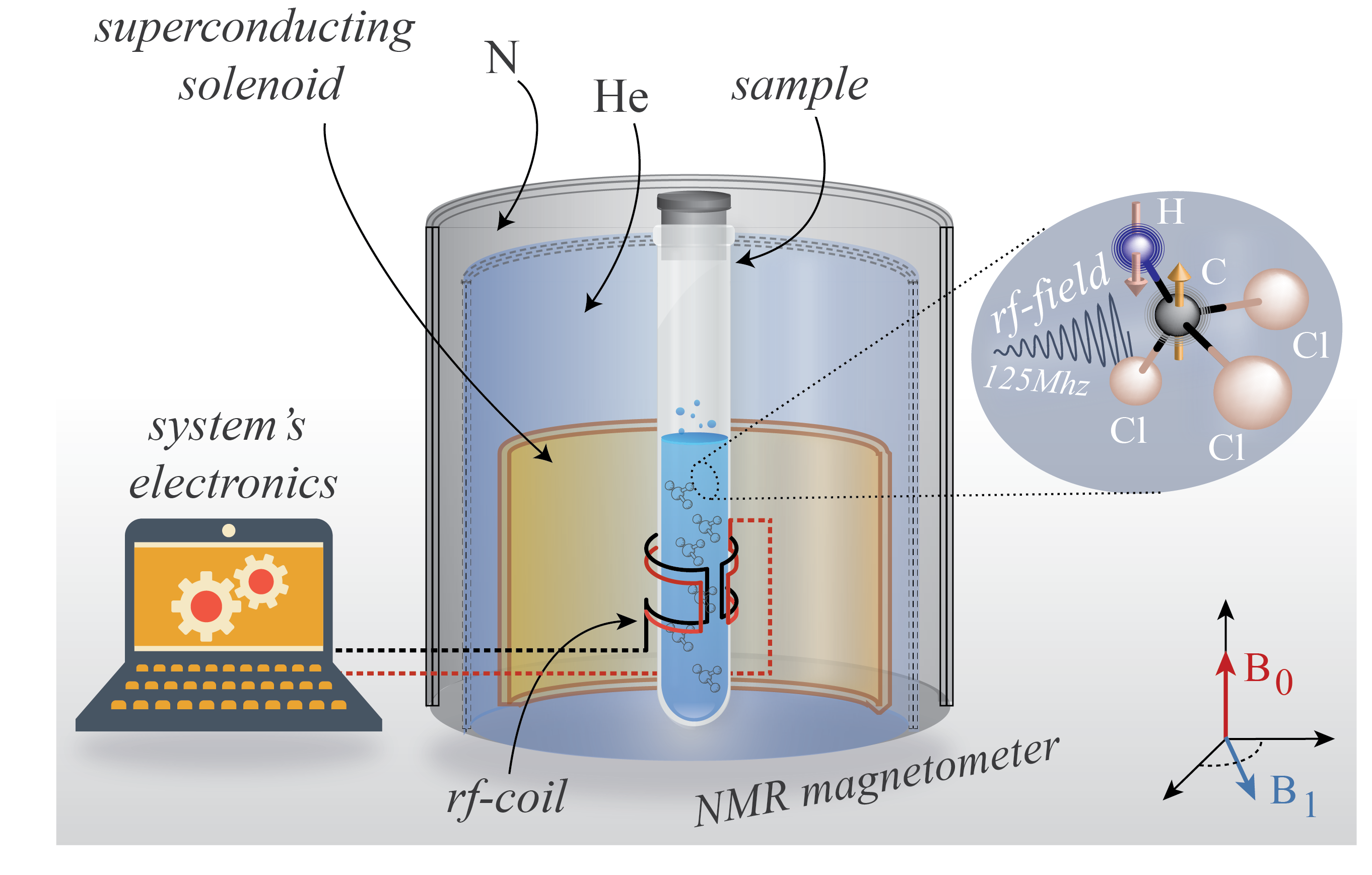}
\caption{ We show a section of the magnetometer employed in our NMR experiment. A superconducting magnet, which produces a high intensity magnetic field ($B_{0}$) in the longitudinal direction,
is immersed in liquid He, surrounded by liquid N in another vacuum
separated chamber, in a thermally shielded vessel. The liquid sample
(inside a 5mm glass tube) is placed at the center of the magnet within
the rf-coil of the probe head. A digital electronic
time-modulated-pulse induces a transverse rf-field ($B_{1}$) that
drives the $^{13}{\rm C}$ nuclear spins out of equilibrium. In the forward (backward) protocol, the rf-coil can
perform (can retrieve) work on (produced by) the nuclear spin sample.
Depending on the speed of the modulation of the driving field, irreversible entropy is produced.
The sketch is not in scale and has been stripped of unnecessary technical details of the 
apparatus.}
\label{fig:setup}
\end{figure} 

We consider a liquid-state sample of chloroform, and encode our system in the qubit embodied by the nuclear spin of $^{13}$C~\cite{Ivan, Vandersypen, Cory}. 
The sample is placed in the presence of a longitudinal static magnetic field (whose direction is taken to be along the positive $z$ axis) with strong intensity, $B_0 \approx 11.75$~T. The $^{13}$C nuclear spin precesses around $B_0$ with Larmor frequency ${\omega_\text{C}}/{2\pi} \approx 125$~MHz. We  control the system magnetization through rf-field pulses in the transverse ($x$ and $y$) direction~\cite{Cory}. The initial thermal state $\rho^\text{eq}_0$ of the $^{13}$C nuclear spin (at inverse temperature $\beta$) is prepared by suitable sequences of transversal radio-frequency (rf)-field and longitudinal field-gradient pulses. We use the value  $k_B T/h = 1.56 \pm 0.07$~kHz (corresponding to an effective temperature of $T\simeq75\pm3$nK) throughout the experiment for the initial  $^{13}$C nuclear spin thermal states. A sketch of the experimental setup is provided in Fig.~\ref{fig:setup}.

 \begin{figure}[t!]
\includegraphics[width=\columnwidth]{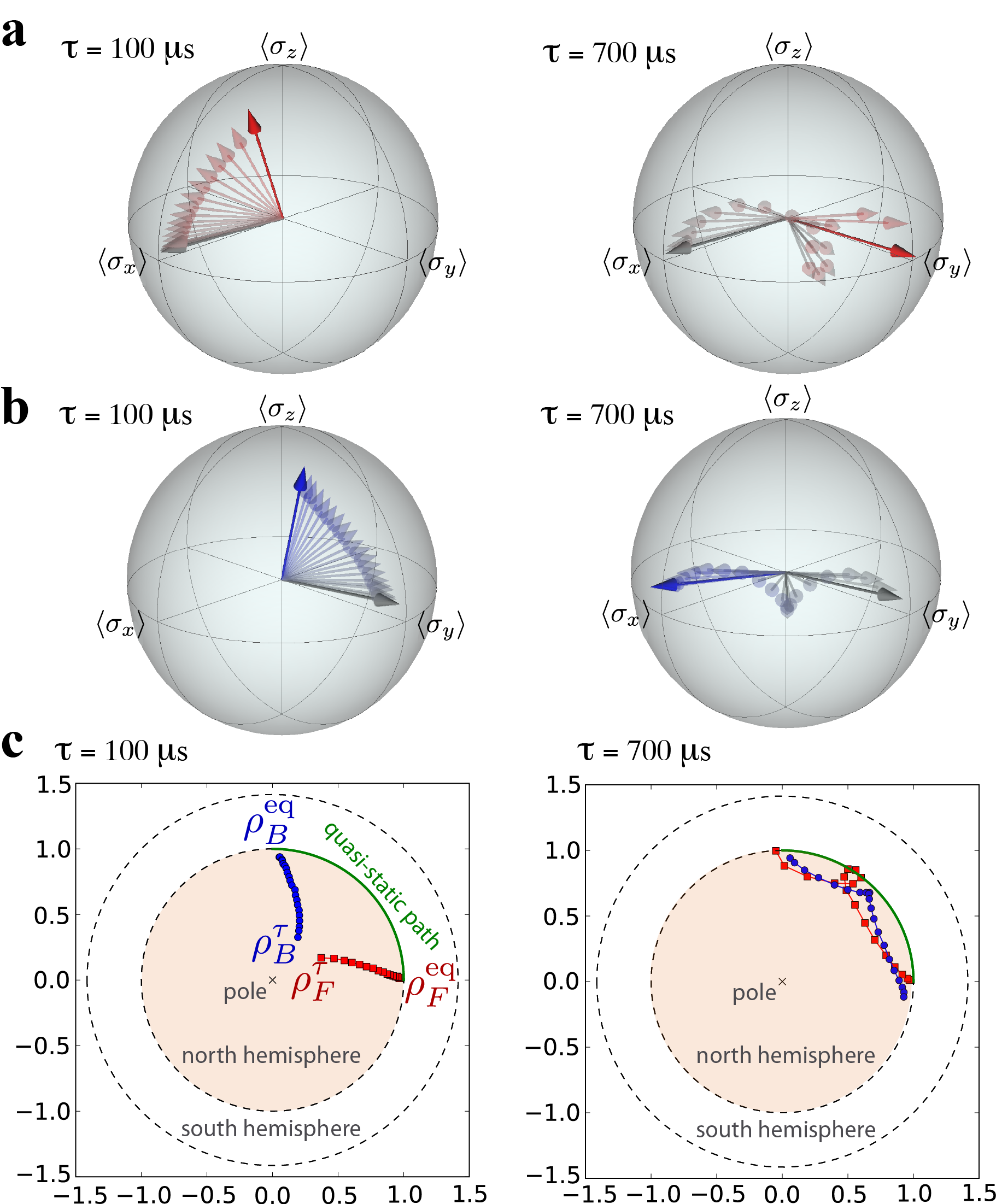}
\caption{ {\bf (a)} [{\bf (b)}] Evolution of the Bloch vector of the forward [backward] spin-1/2 state ${\rho}^\text{F}_t$ [${\rho}^{\text{B}}_{\tau-t}$] during a quench of the transverse magnetic field, obtained via quantum state tomography. A sampling of 21 intermediate steps has been used. The initial magnetization (gray arrow) is parallel to the external rf-field, aligned along positive $x$ [$y$] axis for the forward [backward] process. The final state is represented as a red [blue] arrow. {\bf (c)} Polar projection (indicating only the magnetization direction) of the Bloch sphere with the trajectories of the spin. Green lines represent the path followed in a quasistatic ($\tau\rightarrow \infty$) process.}
\label{fig:mag}
\end{figure}

The system is  driven out of equilibrium to the state $\rho^\text{F}_\tau$ by a fast quench of its Hamiltonian (denoted as ${\cal H}^\text{F}_t$ in this {\it forward} process) lasting a time $\tau$. We experimentally realize this quench by a transverse time-modulated rf field set at the frequency of the nuclear spin. In a rotating frame at the spin Larmor frequency, the Hamiltonian regulating the forward process is    
\begin{align}
{\cal  H}^\text{F}_t &= 2\pi\hbar \nu\left( t \right)
 \left[ \sigma_x^{\text{C}}\cos \phi (t)  + \sigma_y^{\text{C}}\sin \phi (t)   \right]
 \label{eq:quench}
\end{align} 
with ${\sigma}_{x,y,z}^{\text{C}}$ the Pauli spin operators, ${\phi (t)={\pi}t/({2\tau})}$, and ${\nu(t)=\nu_0 \left(1-t/{\tau}\right) + \nu_{\tau} t/{\tau}}$ the (linear) modulation of the rf-field frequency over time $\tau$, from value $\nu_0 = 1.0\text{ kHz}$ to $\nu_{\tau} = 1.8\text{ kHz}$. With these definitions, the initial thermal state of the $^{13}{\rm C}$ system is $\rho^\text{eq}_0=\exp(-\beta{\cal H}^F_0)/Z_0$, where $Z_0$ is the partition function at time $t=0$.

\begin{figure*}[t!]
{\bf (a)}\hskip6cm{\bf (b)}\hskip6cm{\bf (c)}\\
\includegraphics[width=0.66\columnwidth]{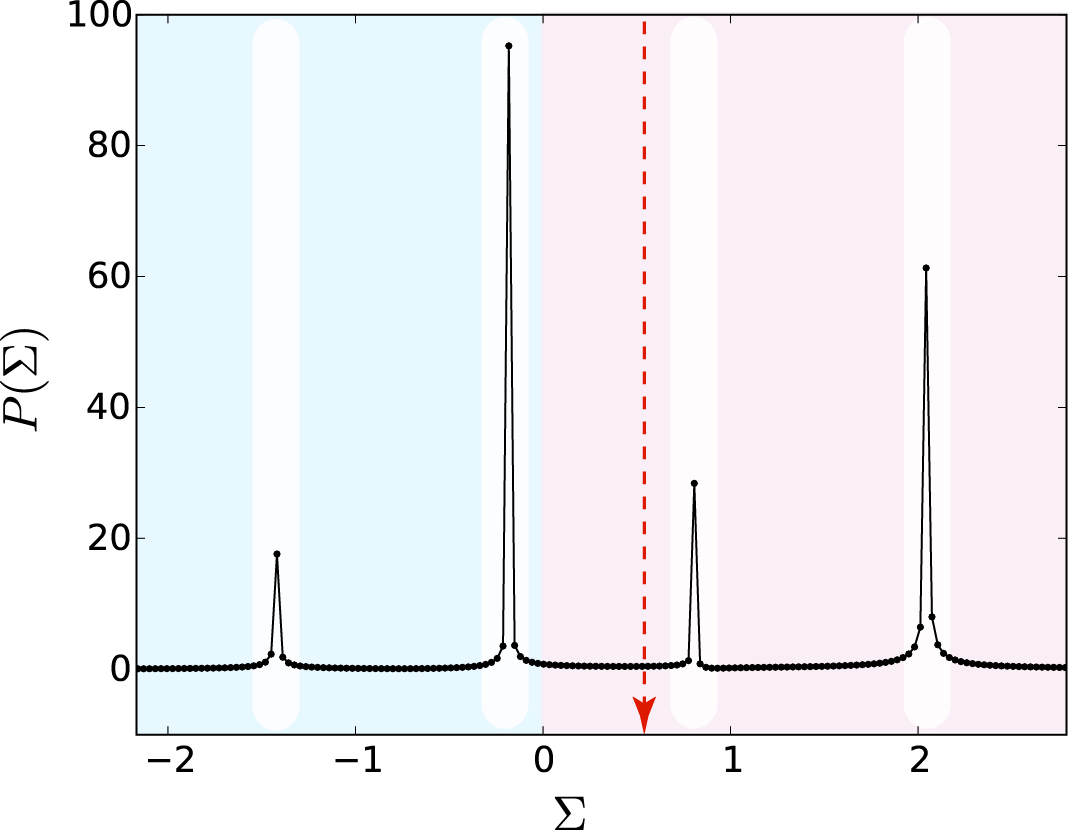}\quad\includegraphics[width=0.66\columnwidth]{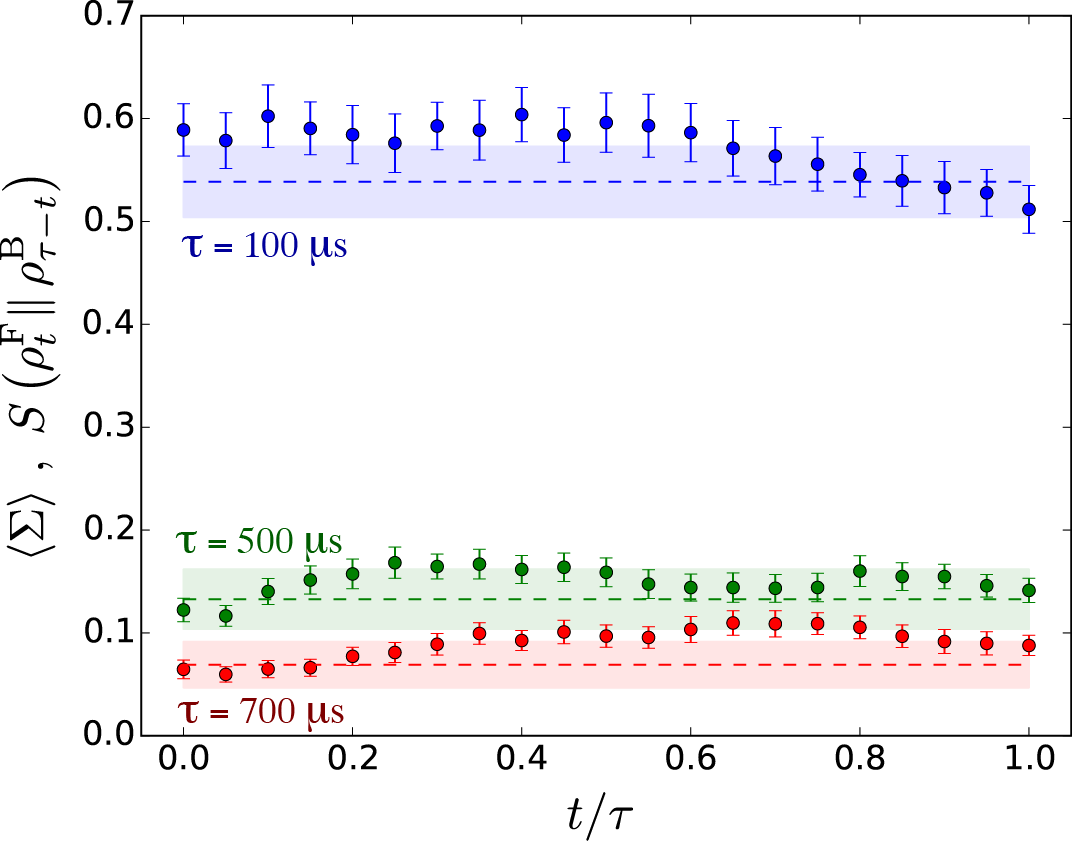}\quad\includegraphics[width=0.66\columnwidth]{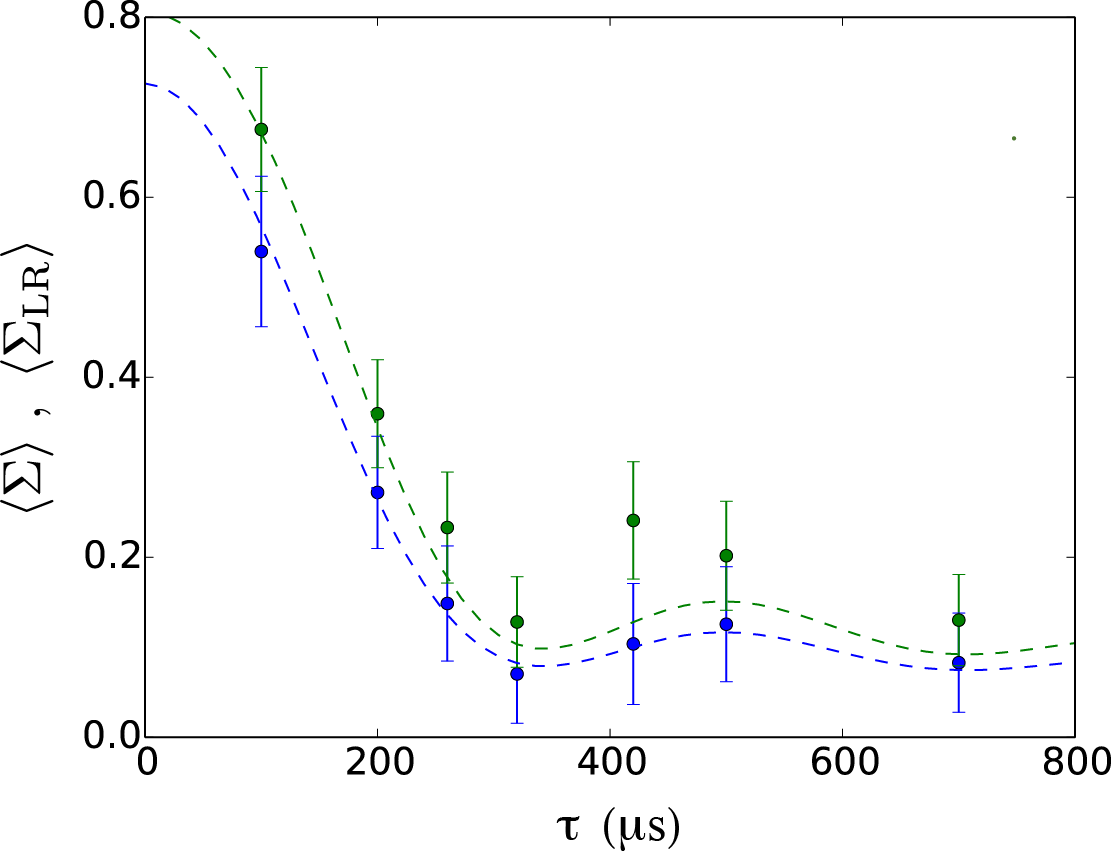}
\caption{ {\bf (a)} Black dots represent the measured negative and positive values of the entropy production $\Sigma$ of  the spin-1/2 system after a quench of the transverse magnetic field of duration $\tau = 100$ $\mu$s. The mean entropy production (red dashed line) is positive, in agreement with the second law. {\bf (b)} Average entropy production $\langle \Sigma\rangle$ (dashed lines) evaluated through the probability distribution $P(\Sigma)$, and Kullback-Leibler divergence $S\left( {\rho}^\text{F}_{t} \Vert \, {\rho}^{\text{B}}_{\tau-t}\right)$ between forward and backward states ${\rho}^\text{F}_t$ and ${\rho}^{\text{B}}_{\tau-t}$ (dots), reconstructed through the tomographic measurements, as a function of time for three different quench durations $\tau = 100~\mu$s (blue), $500~\mu$s (green), and $700~\mu$s (red). Good agreement (within experimental uncertainties represented respectively by error bars and shadowed regions) between the two quantities is observed, quantifying the arrow of time Eq.~(1). {\bf (c)} Mean entropy production $\langle \Sigma\rangle$ (blue dots) and its linear response approximation $\langle\Sigma_\text{LR}\rangle$ (green dots) as a function of the quench time. The difference between the two values, especially for fast quenches (small $\tau$), suggests the systematic deviation of the experiment from the linear response regime. The dashed lines represent the results of numerical simulations.}
\label{fig:entropydist}
\end{figure*} 

In order to reconstruct the work and entropy production statistics of the $^{13}$C quenched dynamics, we use the method proposed in Refs.~\cite{Batalhao,Dorner,Mazzola} and illustrated in detail in the Appendix. We consider an ancillary qubit embodied by the $^1$H nuclear spin of our sample (Larmor frequency ${\omega_\text{H}}/{2\pi} \approx 500$~MHz) and exploit the natural scalar coupling between the $^1$H and $^{13}$C nuclear spins to implement the interferometer needed to reconstruct the statistics of the work done by $^{13}$C following the quench (cf. Appendix). The method assumes a unitary dynamics of both system and ancilla, a condition that is met with excellent accuracy in our experiment. In fact, the spin-lattice relaxation times, measured by inversion recovery pulse sequence, are $({\cal T}_{1}^{\text{H}}, {\cal T}^{\text{C}}_1) \approx (7.36,10.55)$~s. The transverse relaxations, obtained by CPMG pulse sequence, 
have characteristic times $({\cal T}^{\text{H}}_{2},{\cal T}^{\text{C}}_{2})\approx(4.76,0.33)$~s. We thus study processes of maximal duration $\tau\sim10^{-4}$~s and consider total data acquisition times for the reconstruction of the work and entropy statistics within $0.1$~ms and $126$~ms, being smaller than ${\cal T}^{\text{H,C}}_{1}$. This enables us to disregard any energy exchange with the system environment during the quenched dynamics. The effects of the $^{13}$C transverse relaxation (${\cal T}^{\text{C}}_{2}$) can be partially overcome by a refocussing strategy in the reconstruction procedure. 

We implement the {\it backward} process, shown in Fig. 1, by driving the system with the time-reversed Hamiltonian, ${\cal H}^\text{B}_t= {\cal H}^\text{F}_{\tau-t}$, from the equilibrium state, $\rho^\text{eq}_\tau=\exp(-\beta {\cal H}^\text{F}_\tau)/Z_\tau$, that corresponds to the final Hamiltonian ${\cal H}^\text{F}_\tau$ ($Z_t$ here denotes the partition function at time $t$). The intermediate $^{13}$C states during the field quench are ${{\rho}^{\text{F}}_{t}  = {\cal U}_{t} {\rho}^{\text{eq}}_0 {\cal U}_{t}^{\dagger}}$ and ${{\rho}^{\text{B}}_{\tau-t} = {\cal V}_{\tau-t} {\rho}^{\text{eq}}_{\tau} {\cal V}_{\tau-t}^{\dagger}}$, where the evolution operators satisfy the time-dependent Schr\"odinger equations ${d_t {\cal U}_{t} = - i {\cal H}_t^\text{F} {\cal U}_{t}}$ and  ${d_t {\cal V}_{t} = i {\cal H}_{\tau-t}^\text{F} {\cal V}_{t}}$ with the initial conditions ${\cal U}_{0} = {\cal V}_{0} = {\openone}$.

 Work is performed on the system during the forward and backward processes. The corresponding probability distributions $P^{\text{F,B}}(W)$ are related via the Tasaki-Crooks fluctuation relation \cite{Tasaki,Crooks,quantum}
\begin{equation}
\label{eq:Crooks}
 P^{\text{F}}\left( W \right)/P^{\text{B}}\left( -W \right)= e^{\beta \left(W - \Delta F\right)}.
\end{equation}
Equation \eqref{eq:Crooks} characterizes the positive and negative fluctuations of the quantum work $W$ along single realizations. It holds for any
driving protocol, even beyond the linear response regime, and is a generalization of the second law to which it reduces on average, $\langle \Sigma\rangle =\beta(\langle W\rangle-\Delta F)\geq 0$.

We experimentally verify the arrow of time expressed by Eq.~\eqref{eq:rel-entropy} by determining both sides of the equation independently. We first evaluate the Kullback-Leibler relative entropy between forward and backward dynamics by tracking the state of the spin-1/2 at any time $t$ with the help of quantum state tomography  \cite{Ivan}. Figs.~\ref{fig:mag} {\bf (a)}-{\bf (c)} show reconstructed trajectories followed by the Bloch vector, for both forward and backward processes and different quench times. As a second step, we measure the probability distribution $P(\Sigma)$ of the irreversible entropy production  using the Tasaki-Crooks relation Eq.~\eqref{eq:Crooks}. Employing NMR spectroscopy \cite{Ivan} and the method described in Refs. \cite{Batalhao,Dorner,Mazzola} (cf. Appendix for a detailed analysis), we determine the forward and backward work distributions, from which we  extract $\beta$, $W$ and $\Delta F$, and hence the  entropy produced during each process. The measured nonequilibrium entropy distribution is shown in Fig.~\ref{fig:entropydist} {\bf (a)}. It is discrete as expected for a quantum system. We further observe that both positive and negative values occur owing to the stochastic nature of the problem. However, the mean entropy production is positive (red line) in agreement with Clausius inequality $\langle \Sigma \rangle \geq 0$ for an isolated system. We have thus directly tested one of the fundamental expressions of the second law of thermodynamics at the level of an isolated quantum system \cite{cal85}.

A comparison of the  mean entropy production with the Kullback-Leibler relative entropy between forward and backward states is displayed in Fig.~\ref{fig:entropydist} {\bf (b)} as a function of the quench time. We observe good agreement between the two quantities within experimental errors that are due to inhomogeneities in the driving rf-field and non-idealities of the field modulation. These results  provide a first experimental confirmation of Eq.~(1) and the direct verification of the thermodynamic arrow of time in a driven quantum system. {They quantify in a precise manner the intuitive notion that the faster a system is driven away from thermal equilibrium (i.e. the bigger the mean  entropy production or the shorter the driving time $\tau$), the larger the degree of irreversibility, as measured by the relative entropy between a process and its time reversal. }


In the linear response regime \cite{cal85}, Onsager has derived generic expressions for the entropy production which form the backbone of standard non equilibrium thermodynamics. These results are, however,   limited to systems that are driven close to thermal equilibrium. By contrast, Eq.~\eqref{eq:rel-entropy} holds for any driving protocol and thus arbitrarily far from equilibrium. In order to check the general validity of Eq.~\eqref{eq:rel-entropy}, we use the linear response (LR) approximation of the mean work \cite{Liphardt}, $\langle W_{\text{LR}}\rangle = \Delta F+\beta  \Delta W^2/2$, where $\Delta W^2$ is the variance of the work, to obtain the mean entropy production $\langle\Sigma_{\text{LR}}\rangle =  \beta^2 \Delta W^2/2$. Fig.~\ref{fig:entropydist} {\bf (c)} shows the experimental values of $\langle\Sigma \rangle$ and $\langle\Sigma_{\text{LR}}\rangle$ as a function of the quench duration. We note that the measured irreversible entropy production $\langle\Sigma \rangle$ is close yet systematically distinct from its linear response  approximation $\langle\Sigma_{\text{LR}}\rangle$, the difference being more pronounced for fast quenches (small $\tau$), as expected. Fig.~\ref{fig:entropydist} {\bf (c)} thus suggests that the quenches implemented in the experiment are performed somewhat away from the nonlinear response regime. We also mention that we achieve good agreement between experimental data (dots) and numerical simulations (dashed lines).


\noindent{\it Conclusions.--} We have assessed the emergence of the arrow of time in a thermodynamically irreversible process by using the tools provided by the framework of non equilibrium quantum thermodynamics. We have implemented a fast quenched dynamics on an effective qubit in an NMR setting, assessing both the mean entropy produced across the process and the {\it distance} between the state of the system and its reverse version, at all times of the evolution. Let us discuss the physical origin of such time-asymmetry in a closed quantum system. Using an argument put forward by Loschmidt in the classical context, its time evolution should in principle be fully reversible~\cite{leb93}. {How can then a unitary equation, like the Schr\"odinger equation, lead to Eq.~\eqref{eq:rel-entropy} that contains a strictly nonnegative  relative entropy? The answer to this puzzling question lies in the observation that the description of physical processes requires both equations of motion \textit{and} initial conditions~\cite{leb93,jar11}. The choice of an initial thermal equilibrium state singles out a particular value of the entropy,  breaks time-reversal invariance and thus leads to  the arrow of time.} The dynamics can only be fully reversible for a genuine equilibrium process for which the entropy production vanishes at all times. Moreover, issues linked to the "complexity" of the preparation of the initial state to be used in the forward dynamics (or the corresponding one associated with the time-reversed evolution) have to be considered~\cite{lesovik}. By providing an experimental assessment of the microscopic foundation of irreversibility (systematically beyond the linear response regime), our results both elucidate and quantify the physical origin of the arrow of time in the quantum setting of an isolated system.

\section*{Appendix}

\noindent \textbf{1. Initial thermal state preparation}\\
Preparation of the initial state can be done employing spatial averaging techniques, i.e.,
a combination of transverse rf pulses and longitudinal field gradients~\cite{Ivan, Jones}.
Through this standard method, we prepared a joint pseudo pure state equivalent to $\ket{0}_\text{H}\hspace{-0.1cm}\bra{0} \otimes
\rho^{\text{eq}}_{\alpha}$ for the Hydrogen and the Carbon nuclear spins, with $\alpha=0,\tau$
for the forward and backward protocols, respectively.
We are using the logical representation of the ground ($\ket{0}$) and excited ($\ket{1}$) spin states.
The Hydrogen nuclear spin is prepared in the 
ground state while the Carbon is effectively prepared in a thermal diagonal state 
$\rho^{\text{eq}}_{\alpha} = e^{-\beta {\cal H}^\text{F}_{\alpha}} / {Z}_{\alpha}$,
where ${Z}_\alpha=\operatorname{tr} e^{-\beta {\cal H}^\text{F}_{\alpha}}$ is the partition function for ${\cal H}^\text{F}_\alpha$. The initial populations of $\rho^{\text{eq}}_{\alpha}$ in the ${\cal H}^\text{F}_{\alpha}$ basis follow the Gibbs distribution associated with inverse spin (pseudo-)temperature $\beta= (k_B T)^{-1}$.

In all experiments we fixed the spin temperature of the Carbon in the initial Gibbs thermal
distribution, such that ${k_B T/h = 1.56 \pm 0.07 \text{ kHz}}$. The uncertainty in this value is obtained through a
comparison of Carbon populations in a series of independent state preparation 
and quantum state tomography. The trace distance between the experimentally prepared initial state
($\rho^{\text{exp}}$) and the ideal Gibbs state ($\rho^{\text{ideal}}$), 
defined as $\delta \equiv\operatorname{tr} |\rho^{\text{exp}} - \rho^{\text{ideal}} |_1/2$,
is smaller than $0.03$ for all prepared states, in other words a less than $3\%$ chance to discriminate them.  \\

\noindent \textbf{2. Work and irreversible entropy distributions.}\\
\noindent The work performed on or by a closed quantum system undergoing a protocol (dictated by ${\cal H}^{\text{F,B}}_t$)  is a stochastic variable~\cite{Esposito, Campisi}. For a quasi-static and isothermal process, the average work equals the variation in free energy. On the other hand, in a scenario involving a non-quasistatic and driven closed system dynamics (as in our experiment), an amount of work may be done on the system above the variation in free energy. This introduces a \emph{dissipated work}, $\langle W_{\text{diss}} \rangle= \langle W\rangle - \Delta F \ge 0$, inherently linked to the irreversible nature of the process. Such irreversibility is associated to an entropy production \cite{jar11, def10}, 
\begin{equation}
\Sigma = \beta \left( W - \Delta F \right) \, .
\label{eq:defentropyprod}
\end{equation}
which is also a stochastic variable that depends on the work distribution, the net change in the system free energy, and the inverse temperature. So the entropy production probability distribution, $P(\Sigma)$, is built upon the experimental assessment of all the mentioned variables. 

\begin{figure*}[t]
 \includegraphics[scale=0.38]{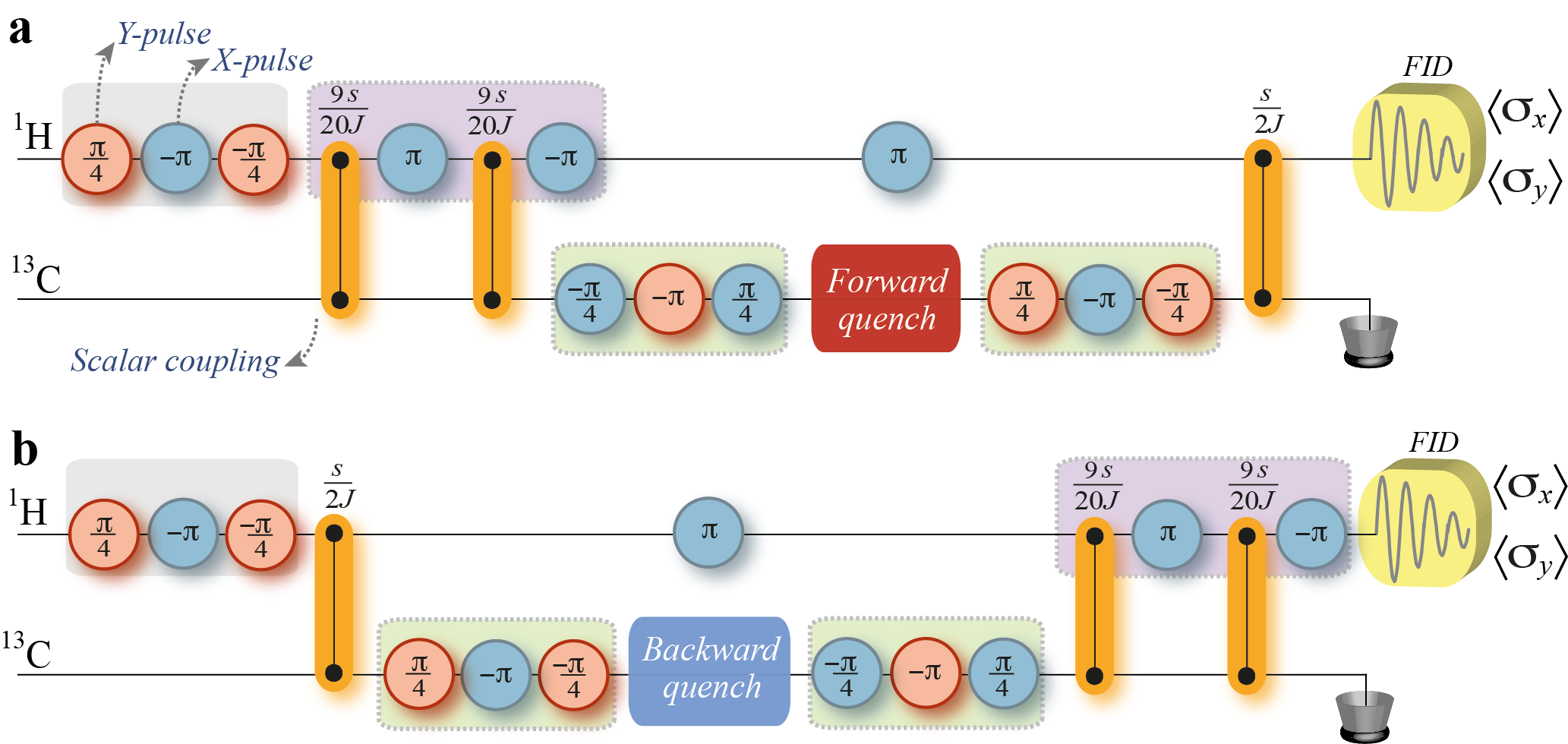}
 \caption{{NMR pulse-sequence for the reconstruction of the work characteristic functions
 $\chi^{\text{F}}\left(u\right)$ and $\chi^{\text{B}}\left(u\right)$}.
 \textbf{a}, Sequence for the forward process ($\alpha = 0$).
 \textbf{b}, Sequence for the backward process ($\alpha = \tau$). 
 We start from the pseudo-pure state $\ket{0}_\text{H}\hspace{-0.1cm}\bra{0} \otimes \rho_\alpha^{\text{eq}}$.
The blue (red) circles represent transverse rf-pulses in the $x$ ($y$) direction that produce rotations 
by the displayed angle. The orange connections represent free evolutions under the scalar interaction, 
${\cal H}_J=2\pi J \sigma^\text{H}_ z\sigma^\text{C}_ z$  (with $J \approx 215.1$~Hz), during the time displayed above each connection. This time-length of the coupling is varied as function of $s$, which is related to the conjugate variable $u$ in Eq.~(\ref{eq:charfunc}) as $s=2\pi \nu_0 u$.}
\label{fig:pulsesequence}
\end{figure*}

An important step in our experiment is the characterization of the probability distribution
of work fluctuations during the quenched dynamics. Such a distribution is obtained using a Ramsey-like interferometric 
method theoretically proposed in refs.~\citep{Dorner,Mazzola} and experimentally implemented 
to verify the fluctuation relations in a quantum scenario~\citep{Batalhao}. This approach is based on the interferometric
measurement of the characteristic function of the work distribution~\cite{Talkner}, which can be written 
for the forward process ($\text{F}$) as   
\begin{equation}
\label{eq:charfunc}
\chi^{\text{F}}(u) =\sum_{m,n}p^{\text{eq},0}_{n}p^{\text{F},\tau}_{m\mid n}e^{iu(\overline{\epsilon}_{m}-\epsilon_{n})} \, , 
\end{equation}
where $p^{\text{eq},0}_{n}=e^{-\beta\epsilon_n}/{Z}_0$ is the initial occupation probability 
for the $n$-th eigenstate ($\ket{n_0}$) of ${\cal H}_0^\text{F}$ with energy $\epsilon_n$, 
while $p^{\text{F},\tau}_{m\mid n}=\left|\langle m_\tau\vert{\cal U}_\tau\vert n_0\right\rangle|^2$ is the conditional 
transition probability to find the system in the $m$-th eigenstate ($\ket{m_\tau}$) of ${\cal H}_\tau^\text{F}$
(with energy $\overline\epsilon_m$) if it was previously in $\ket{n_0}$ at $t=0$. 

The characteristic function can also be written as 
\begin{equation}
\label{eq:charfunctrace}
\chi^{\text{F}}(u) =\operatorname{tr}[ ({\cal U}_\tau e^{-iu{\cal H}_0^\text{F}}) {\rho}_0^\text{eq}
( e^{-iu{\cal H}_\tau^\text{F}}{\cal U}_\tau)^{\dagger}] \, .
\end{equation}
The main idea of the protocol to reconstruct $\chi^{\text{F}}\left(u\right)$ is to estimate the trace in 
Eq.~(\ref{eq:charfunctrace}) via an ancillary system~\citep{Dorner,Mazzola},
the Hydrogen nuclear spin in the present experiment. The NMR pulse sequence employed to perform 
such a task is depicted in Fig.~\ref{fig:pulsesequence}. The ancillary system, initially prepared in the ground 
pseudo-pure state, is rotated to an equal-weighted superposition
(equivalent to $\left(\left|0\right\rangle_{\text{H}} + \left|1\right\rangle_{\text{H}}\right)/\sqrt{2}$) in the
grey box of Fig.~\ref{fig:pulsesequence}. The Carbon nuclear spin is the driven system of our 
interest. For technical reasons (see Sect.~1 above), the $^{13}$C spin starts from a mixture of computational-basis ($\sigma_z^C$-basis) states, with 
occupation probabilities given by $p_n^{\text{eq},0}$ (or by $p_n^{\text{eq},\tau}$ in backward mode). In fact, the state basis will be further rotated to the initial Hamiltonian one. Controlled-unitary operations
are performed using the natural scalar coupling interaction 
${\cal H}_J = 2\pi J \sigma_z^\text{H} \sigma_z^\text{C}$, which is proportional 
to $\sigma_z^C$, while the initial and final Hamiltonians ${\cal H}_0^\text{F}$ 
and ${\cal H}_\tau^\text{F}$ in the forward protocol are proportional to $\sigma_y^\text{C}$ and $\sigma_x^\text{C}$, 
respectively.  This is compensated by the rotations inside the green boxes of Fig.~\ref{fig:pulsesequence}. Such rotations also account for the fact that the initial state was prepared in the $\sigma_z^\text{C}$-basis. The same reasoning is applied in the backward case.
The purple boxes in Fig.~\ref{fig:pulsesequence} are a refocus strategy to mitigate the effects
of the transverse relaxation.
The quenched dynamics is a consequence of a suitable modulation (amplitude and phase) of a transverse rf field, 
which can be described by the Hamiltonian in Eq.~(1)
for the forward protocol. The final step of the algorithm is the measurement of the free induced decay (FID) 
signal of the Hydrogen nuclear spin. 
From this signal one can obtain the transverse magnetization, where the characteristic function is encoded
as $\chi^{\text{F},\text{B}} \left(u\right) = 2\left\langle \sigma_x^{\text{H}} \right\rangle
+2i\left\langle \sigma_y^{\text{H}} \right\rangle$. Application of an inverse Fourier transform allows us to
obtain the work distribution for the quenched dynamics.

\begin{figure}[t]
 \includegraphics[width=1.0\columnwidth]{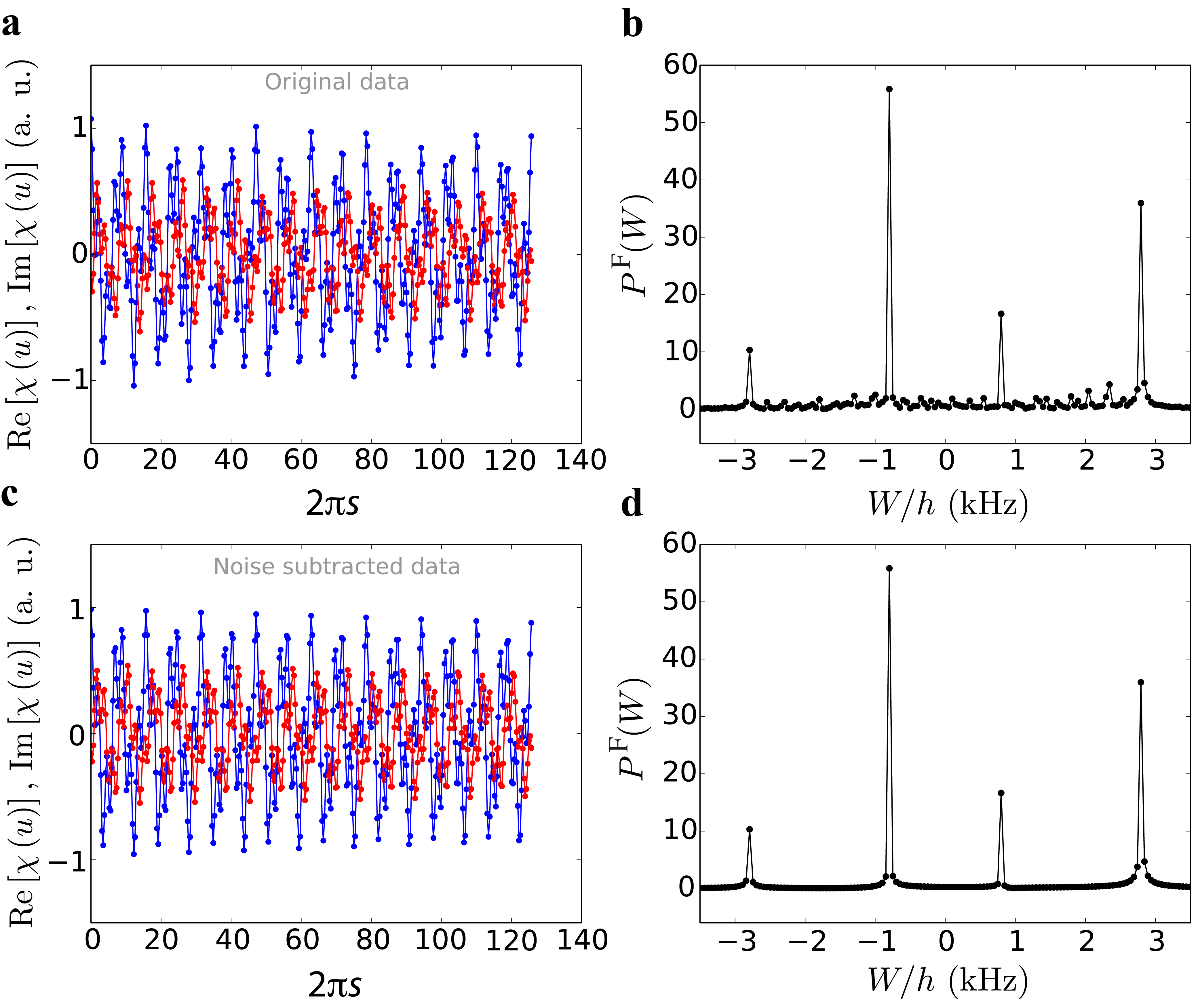}
 \caption{{Typical output of the interferometric circuit shown in Fig.~\ref{fig:pulsesequence}, 
 for a forward quench of time-length $\tau = 100 \, \mu\text{s}$}. \textbf{a}, Blue and red symbols (lines) 
 correspond to the real and imaginary parts of the characteristic function $\chi^{\text{F}}\left(u\right)$
 (where $u =(2\pi \nu_0)^{-1}s$), i.e., the experimentally measured 
 $x$ and $y$ components of the $^1$H transverse magnetization displayed in arbitrary units (a. u.),
 as function of the adimensional parameter $s$.
 \textbf{b}, Work distribution $P^\text{F}(W)$ obtained from the inverse Fourier transform of the the characteristic
 function. \textbf{c} and \textbf{d}, Same as panels a and b, but after subtracting the noise in the Fourier analyses.
}
\label{fig:examplemag}
\end{figure}

We have measured several experimental configurations, keeping fixed the initial spin temperature
(given by the weights $p_n^{\text{eq},0(\tau)}$ of the initial Carbon population), and varying the quench
type (forward or backward) and the quench duration. For each configuration, the interaction time $s$ 
of the free evolution under the scalar coupling in Fig.~\ref{fig:pulsesequence} was varied through $360$ 
equally-spaced values; each realisation corresponds to an independent experiment with an average over a 
set of identical and independent molecules.

A typical output of the characteristic function reconstruction algorithm is shown in Fig.~\ref{fig:examplemag} {\bf (a)}, 
where each data point corresponds to an independent experiment for each scalar interaction parameter $s$. 
The inverse Fourier transform of the transverse magnetization (the work distribution $P^{\text{F}}(W)$) is displayed
in Fig.~\ref{fig:examplemag} {\bf (b)}. From Eq.~(\ref{eq:charfunc}),
we would expect to observe four peaks, since we have a quenched two-level system. Furthermore, we expect the peaks
to be located around $\{\pm(\nu_0 \pm \nu_\tau)\}$,
which correspond to the frequencies $\left\{-2.8,-0.8,+0.8,+2.8\right\}$~kHz considering the parameters used in our setup
(${\nu_0 = 1.0\text{ kHz}}$ and ${\nu_{\tau} = 1.8\text{ kHz}}$).
The peaks in Fig.~\ref{fig:examplemag} {\bf (b)} are proportional to the probabilities $p^{\text{eq,0}}_1 p^{\text{F},\tau}_{0|1}$, 
$p^{\text{eq,0}}_0 p^{\text{F},\tau}_{0|0}$, $p^{\text{eq,0}}_1 p^{\text{F},\tau}_{1|1}$, and
$p^{\text{eq,0}}_0 p^{\text{F},\tau}_{1|0}$, from left to right, respectively.

A least-squares method can be applied to estimate the amplitudes of the four peaks in Fig.~\ref{fig:examplemag} {\bf (b)},
which can be identified with the terms $p_n^{\text{eq},0} p_{m\mid n}^{\text{F},\tau}$ of Eq.~(\ref{eq:charfunc}). 
We can also use this data to filter the noise frequencies in the original characteristic function
(cf. Fig.~\ref{fig:examplemag} {\bf (a)}] disregarding contributions of other frequencies. 
The noise sources are non-idealities in the refocus sequence of the scalar interaction 
(purple boxes in Fig.~\ref{fig:pulsesequence}).
The noise-subtracted characteristic function
and work probability distribution are shown in Figs. \ref{fig:examplemag}c and \ref{fig:examplemag}d,
respectively. The typical noise amplitude was added to the error bar in the estimation of 
the total probabilities $p_n^{\text{eq},0} p_{m\mid n}^{\text{F},\tau}$. \\

\begin{figure}
 \includegraphics[width=1.03\columnwidth]{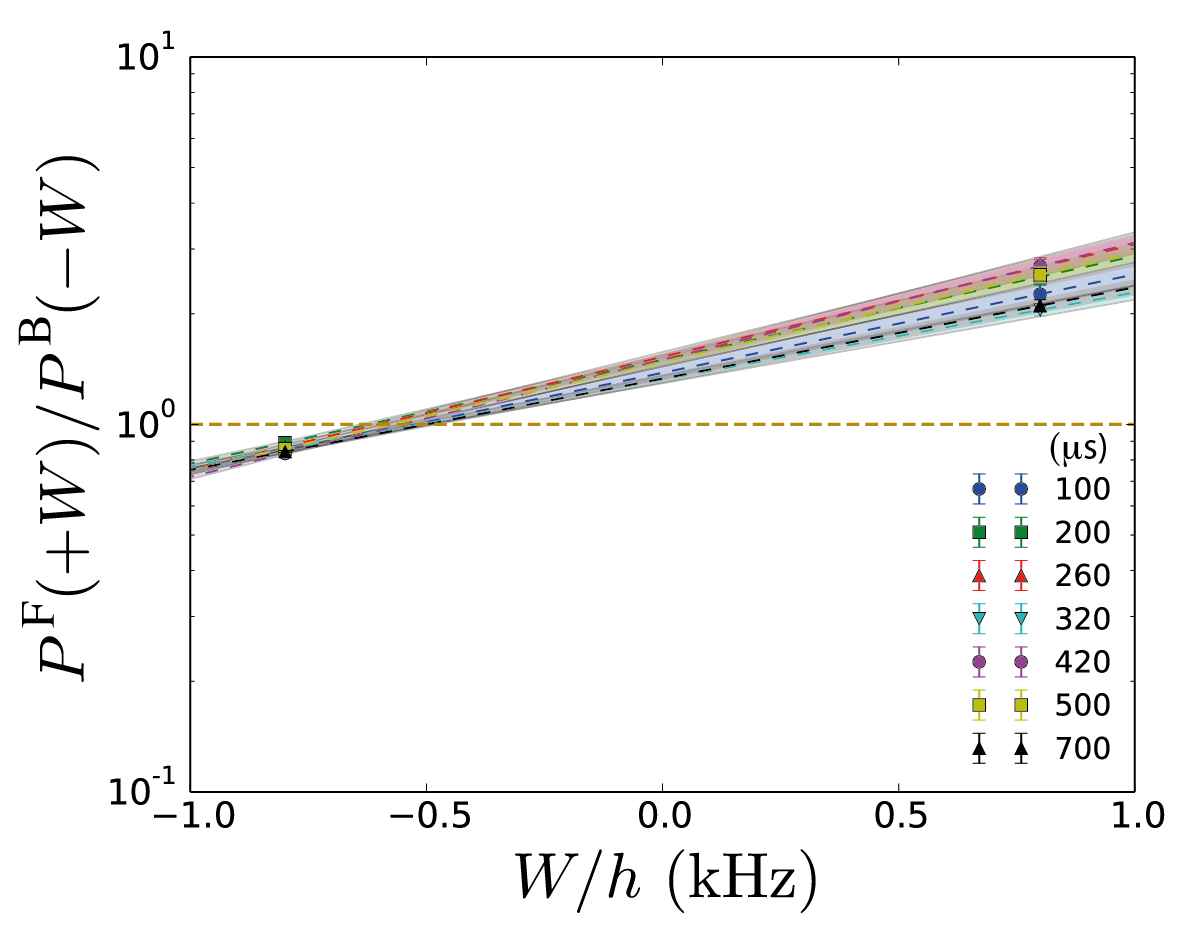}
 \caption{{Estimation of the net difference in free energy}. 
 Crooks relation ${P^\text{F}\left(+W\right)}/{P^\text{B}\left(-W\right)}$ (in logarithmic scale) 
 for different quench time-lengths. Symbols represent experimental data; the dashed line and shaded regions represent 
 the best linear fit and its estimated uncertainty, respectively. The net difference in the free energy,
 $\Delta F$, can be obtained from the interception of the curves and the yellow dashed line 
 (where $P^{\text{F}}(W)=P^{\text{B}}(-W)$ and $\beta W = \beta \Delta F$).
}
\label{fig:crooks}
\end{figure}

\noindent \textbf{3. Estimation of the net difference in free energy.}\\
The Crooks relation can be used to experimentally estimate the free energy difference in 
our nonequilibrium dynamics. We have performed a set of experiments for 
different quenches with durations of $\tau = 100$, $200$, $260$, $320$, $420$, $500$, and $700$ $\mu$s.
For each quench duration, we performed both the forward and backward protocols (as described in Fig.~\ref{fig:pulsesequence}). The ratio between the forward and backward work distributions are plotted in a logarithmic scale in 
Fig.~\ref{fig:crooks} for the different quenches, where we found the best fitting line to the equation (which represents the Crooks relation):
\begin{equation}
 \ln\frac{P^\text{F}\left(+W\right)}{P^\text{B}\left(-W\right)} = \beta (W - \Delta F) \, .
\end{equation}

The data points in Fig.~\ref{fig:crooks} represent the ratio ${P^\text{F}\left(+W\right)}/{P^\text{B}\left(-W\right)}$
only for the inner peaks of the work distributions. Since these points have themselves an uncertainty (from the Fourier analyses),
the linear adjustment also has an uncertainty, which is represented by shaded regions in Fig.~\ref{fig:crooks}.
The angular coefficient of the linear fitting is the inverse spin temperature $\beta$ while the linear coefficient is the product $\beta \Delta F$. \\

\noindent \textbf{4. Entropy production expanded in terms of the cumulants of the work distribution.}\\
\noindent The Jarzynski equality allows one to write the free-energy difference as $\Delta F = - \beta^{-1} \ln \left\langle e^{-\beta W}\right\rangle$, which can be plugged in Eq. (\ref{eq:defentropyprod}) resulting in the following expansion 
\begin{equation}
\left\langle \Sigma \right\rangle = \sum_{n\ge 2} \frac{(-1)^n}{n!} \kappa_n(\beta) \beta^{n} \, ,
\end{equation}
with $\{\kappa_n (\beta)\}$ the set of cumulants of the work distribution. The first two cumulants are the work average, $\kappa_1 = \left\langle W \right\rangle$, and the work-distribution variance, $\kappa_2 = \left(\Delta W\right)^2 = \left\langle \left(W-\left\langle W \right\rangle \right)^2 \right\rangle$, respectively. 
As a first approximation, the mean produced entropy is proportional to the work-distribution variance. If the system presents a linear response to the driving protocol, the relation $\left\langle\Sigma_{LR}\right\rangle =  \beta^2 \left(\Delta W\right)^2/2$ turns out to be a very good approximation. Moreover, at macroscopic scales, the total work involves, in general, a sum of many independent random variables. In this case, the Central Limit Theorem implies that the total work distribution converges to a  Gaussian one, for which all cumulants of order higher than two vanish, making the above approximation exact.  

In our experiment, we are dealing with the microscopic behaviour of a single quantum system driven by a quenched evolution, which leads to the entropy production distribution (and also the work distribution) notably non-Gaussian, as can be observed in Fig.~4 {\bf (a)} of the main text and also in Fig.~\ref{fig:examplemag}. Furthermore our spin system responds in a non-linear way to the field quench as noted in Fig.~3\textbf{ (a), (b)} of the main text. The system magnetization acquires a longitudinal component during the dynamics driven by a fast time-varying Hamiltonian in the transverse direction. These facts introduce a finite deviation from the expectations of linear response theory  as observed in Fig.~4 {\bf (c)} of the main text. We can also note in this figure that for a ``long duration'' quench the mean entropy produced tend to a finite value, since we have a closed system dynamics (it has no thermalisation along the quench protocol). 

The numerical simulations curves, dashed lines plotted in Fig.~4 {\bf (c)} of the main text, were obtained through a numerical integration of the conditional transition probabilities, $p^{\text{F},\tau}_{m\mid n}=\left|\langle m_\tau\vert{\cal U}_\tau\vert n_0\right\rangle|^2$ (employing a code written in Python with the aid of the QuTip toolbox~\cite{qutip}), from which we can compute $P(W)$, $P(\Sigma)$, $\left\langle\Sigma\right\rangle$, and $\left\langle\Sigma_{LR}\right\rangle$. \\

\noindent \textbf{5. Derivation of the microscopic expression for the entropy production (Eq.~(1) of the main text).}\\
\noindent For the sake of completeness we will discuss here a derivation of Eq.~(1) in the main text, which will also be very instructive for the purpose of the present article.
Let us consider the scenario described in Fig.~2 of the main text.
The mean entropy production ${\left\langle \Sigma \right\rangle
= \beta\left\langle W \right\rangle - \beta \Delta F}$ in the forward protocol can be written as
\begin{equation}
\left\langle \Sigma\right\rangle 
=\beta\operatorname{tr}\left({\cal H}_{\tau}^\text{F}{\rho}^\text{F}_{\tau}\right)
-\beta\operatorname{tr}\left({\cal H}_0^\text{F}{\rho}^{\text{eq}}_0\right)
- \beta \Delta F \, ,
\label{eq:meanentropy}
\end{equation}
where ${\Delta F =\beta^{-1} \ln\left({Z}_0/{Z}_\tau\right)}$ is the free energy
difference between the equilibrium states ${{\rho}_0^{\text{eq}}
= e^{-\beta{\cal H}_0^\text{F}}/{Z}_0}$ and ${{\rho}_\tau^{\text{eq}}
= e^{-\beta{\cal H}_\tau^\text{F}}/{Z}_\tau}$,
at constant inverse temperature $\beta$.
We can identify
\begin{equation}
{\cal H}_t^\text{F} = -\beta^{-1}\ln {\rho}^{\text{eq}}_t 
- \openone \beta^{-1} \ln{Z}_t \; ,
\label{eq:identification}
\end{equation}
for $t=0$ and $t=\tau$. Using the identification (\ref{eq:identification}) in Eq.  (\ref{eq:meanentropy}), we obtain 
${ \left\langle \Sigma\right\rangle 
= - \operatorname{tr}\left({\rho}^\text{F}_{\tau} \ln{\rho}^{\text{eq}}_{\tau} \right)
+ \operatorname{tr}\left({\rho}^{\text{eq}}_0 \ln{\rho}^{\text{eq}}_0 \right)}$,
which can be written as a sum of the Kullback-Leibler divergence
and a variation in von Neumann entropy
[${{\cal S}_{\text{vN}}\left({\rho}\right) = -\operatorname{tr}(
{\rho} \ln {\rho})}$],
\begin{equation}
 \left\langle \Sigma \right\rangle 
= {\cal S}\left( {\rho}^\text{F}_{\tau} \Vert {\rho}^{\text{eq}}_{\tau} \right)
+ {\cal S}_{\text{vN}}\left( {\rho}^\text{F}_{\tau} \right)
- {\cal S}_{\text{vN}}\left( {\rho}^{\text{eq}}_{0} \right) \, .
\end{equation}
This derivation did not assume any particular relationship between 
${\rho}^{\text{eq}}_{0}$ and ${\rho}^\text{F}_{\tau}$. The assumption of a closed, unitary
dynamics implies that von Neumann entropies of initial and final states are equal, 
and the irreversible entropy produced during the quench is exactly given 
by the Kullback-Leibler divergence of the final nonequilibrium state and 
the equilibrium state of the final Hamiltonian.

In addition, the assumption of unitary dynamics allows us to prove that 
the Kullback-Leibler divergence between the forward and backward dynamics 
is constant during the quench, i.e.,
${{\cal S}\left( {\rho}^\text{F}_{\tau} \Vert \, {\rho}^{\text{eq}}_{\tau} \right) 
= {\cal S}\left( {\rho}^\text{F}_{t} \Vert \, {\rho}^\text{B}_{\tau-t} \right)}$,
with 
\begin{equation}
{\cal S}\left( {\rho}^\text{F}_{t} \Vert \, {\rho}^\text{B}_{\tau-t} \right)
= \operatorname{tr}\left({\rho}^\text{F}_{t} \ln {\rho}^\text{F}_{t} \right)
- \operatorname{tr}\left( {\rho}^\text{F}_{t}  \ln {\rho}^\text{B}_{\tau-t} \right) \; .
\label{eq:kullback}
\end{equation}
We note that the intermediate states during the quench are ${{\rho}^\text{F}_{t}  = {\cal U}_{t} {\rho}^{\text{eq}}_0 {\cal U}_{t}^{\dagger}}$ and 
 ${{\rho}^\text{B}_{\tau-t} = {\cal V}_{\tau-t} {\rho}^{\text{eq}}_{\tau} {\cal V}_{\tau-t}^{\dagger}}$ (for the forward and backward protocol, respectively), where the evolution operators satisfy
 ${{d_t} {\cal U}_{t} = - i {\cal H}_t {\cal U}_{t}}$ and 
 ${{d_t} {\cal V}_{t} = i {\cal H}_{\tau-t} {\cal V}_{t}}$
(with the initial conditions ${\cal U}_{0} = {\cal V}_{0} = {\openone}$).
A simple change of variables leads to 
${{d_t} {\cal V}_{\tau-t} = - i {\cal H}_t {\cal V}_{\tau-t}}$,
which is the same equation as that for ${\cal U}_{t}$, but with a different initial condition,
namely that ${\cal V}_{\tau-t} \mid_{t=\tau} = {\openone}$.
By inspection of the structure of the Schr\"odinger equation, we are allowed to right-multiply
one solution by any constant operator to obtain another solution for a different initial condition.
Doing that, we obtain a relation between the forward and backward dynamics as
\begin{equation}
{\cal V}_{\tau-t} = {\cal U}_{t} {\cal U}^\dagger_{\tau} \, .
\label{eq:backunitary}
\end{equation}
To prove Eq. (1) of the main text, we shall investigate the time evolution of both terms in Eq.~(\ref{eq:kullback}).
The term ${\operatorname{tr}\left({\rho}^\text{F}_{t} \ln {\rho}^\text{F}_{t} \right)}$ is constant
because of the unitary evolution. It remains to show that the term 
${\operatorname{tr}\left( {\rho}^\text{F}_{t}  \ln {\rho}^\text{B}_{\tau-t} \right)}$ is also constant, which can be proven as follows
\begin{align}
\operatorname{tr}\left[ {\rho}^\text{F}_{t}  \ln {\rho}^\text{B}_{\tau-t} \right] &=
\operatorname{tr}\left[{ \cal U}_{t} {\rho}^{\text{eq}}_0 {\cal U}_{t}^{\dagger} 
\ln\left( {\cal V}_{\tau-t} {\rho}^{\text{eq}}_\tau {\cal V}_{\tau-t}^{\dagger}\right) \right]
\nonumber \\ &=
\operatorname{tr}\left[  {\rho}^{\text{eq}}_0 {\cal U}_{t}^{\dagger} 
{\cal V}_{\tau-t} \ln\left({\rho}^{\text{eq}}_\tau\right) {\cal V}_{\tau-t}^{\dagger}
{ \cal U}_{t}\right]
\nonumber \\ &=
\operatorname{tr}\left[  {\rho}^{\text{eq}}_0 {\cal U}_{\tau}^{\dagger} 
\ln\left({\rho}^{\text{eq}}_\tau\right) { \cal U}_{\tau}\right]
\nonumber \\ &=
\operatorname{tr}\left[ {\rho}^\text{F}_\tau \ln\left({\rho}^{\text{eq}}_\tau\right) \right] \, ,
 \end{align}
where Eq.~(\ref{eq:backunitary}) was used in the last-but-one equality. This shows that 
${\cal S}\left( {\rho}^\text{F}_{t} \Vert \, {\rho}^\text{B}_{\tau-t} \right)$ is 
time invariant. In this way we have verified Eq. (1) in the main text.

For a prior derivation of this result, employing different arguments and methods, see Ref.~\citep{par09}.

\noindent
{\it Acknowledgments.--}We thank K. Micadei for valuable discussions. We acknowledge financial support from CNPq, CAPES, FAPERJ, and FAPESP. MP is supported by the John Templeton Foundation (Grant ID 43467) and the CNPq ``Ci\^{e}ncia sem Fronteiras'' programme through the ``Pesquisador Visitante Especial'' initiative (Grant Nr. 401265/2012-9). MP and EL acknowledge the EU Collaborative Project TherMiQ (Grant Agreement 618074) and the COST Action MP1209. RMS acknowledges the Royal Society and Newton Fund through the Newton Advanced Fellowship scheme (reference n. R1660101). This work was performed as part of the Brazilian National Institute of Science and Technology for Quantum Information (INCT-IQ).

\end{document}